# A Big Data Analytical Framework For Portfolio Optimization

# A Big Data Analytical Framework For Portfolio Optimization
Dhanya Jothimani, Ravi Shankar and Surendra S. Yadav
Department of Management Studies, Indian Institute of Technology Delhi
{dhanya.jothimani, ravi1, ssyadav}@dms.iitd.ac.in

**Abstract.** With the advent of Web 2.0, various types of data are being produced every day. This has led to the revolution of big data. Huge amount of structured and unstructured data are produced in financial markets. Processing these data could help an investor to make an informed investment decision. In this paper, a framework has been developed to incorporate both structured and unstructured data for portfolio optimization. Portfolio optimization consists of three processes: *Asset selection, Asset weighting and Asset management*. This framework proposes to achieve the first two processes using a 5-stage methodology. The stages include shortlisting stocks using Data Envelopment Analysis (DEA), incorporation of the qualitative factors using text mining, stock clustering, stock ranking and optimizing the portfolio using heuristics. This framework would help the investors to select appropriate assets to make portfolio, invest in them to minimize the risk and maximize the return and monitor their performance.

## 1. Introduction

Big data has gained popularity in the recent years and has become a buzz word. However, it is time to understand what "Big Data" is. Big Data refers to large data being generated continuously in the form of unstructured data produced by heterogeneous group of applications from social network to scientific computing applications, and so forth. The dataset ranges from a few hundred gigabytes to terabytes that is beyond the capacity of existing data management tools that can capture, store, manage and analyze [1]. Big data is characterized by the following dimensions: *Volume, Velocity, Variety, Variability, Complexity and Low Value Density* [2]. Various types of data, namely, structured, semi-structured and unstructured data are produced with the advent of Web 2.0 technology [3].

In the era of big data, one of the major challenges for the capital market firms is to handle the velocity at which data is being generated, considering the production of unstructured data in financial services like investment banking. New York Stock Exchange (NYSE) and NASDAQ have employed big data applications like IBM Netezza to store and process big data from various sources. Capital market firms use big data technologies to mitigate risks (fraud mitigation, on-demand enterprise management), regulation, trading analytics (High Frequency Trading, Pre-trading decision), and data tagging [4].

The firms adopting big data technologies and predictive analytics have an edge over other firms in the uncertain market conditions [5]. In finance, portfolio is defined as the collection of assets. Assets range from stocks and bonds to real estate. With the seminal work of Markowitz [6], portfolio optimization has been a topic of research. Portfolio optimization is the investment decision-making process to hold a set of financial assets to meet various criteria of the investors. In general terms, the criteria are maximizing return and minimizing risk. In this paper, the scope of the work is limited to stock analysis. Portfolio optimization consists of three major steps: asset selection, asset weighting, and asset management. In this paper, a framework is proposed to integrate unstructured and structured data to make an informed investment decision.

The design of the paper is as follows. Section 2 discusses the proposed framework for portfolio optimization, followed by the expected outcomes in Section 3. Conclusions are presented in Section 4.

**2. Proposed Framework**

The proposed framework for portfolio optimization can be explained using 5-step process: *(a) Data Envelopment Analysis (DEA), (b) Validation of selected stocks, (c) Stock clustering, (d) Stock ranking, and (e) Optimization*. All listed firms at a particular stock exchange are considered as the initial input to the framework and the output would be a set of stocks that would maximize the return and minimize the risk. The abstract framework for portfolio optimization is shown in figure 1. DEA is used to narrow the sample space of firms by identifying the efficient firms. In order to validate these firms as potential candidates for portfolio optimization, the latest information about the company is retrieved and processed from online news articles and tweets using text mining to the sentiments about the company in current context. The validated efficient firms are clustered into different groups to aid the diversification of portfolio. This is further followed by ranking of the stocks within each cluster and followed by asset weighting using optimization algorithms. Each process is explained from section 2.1 to section 2.5.

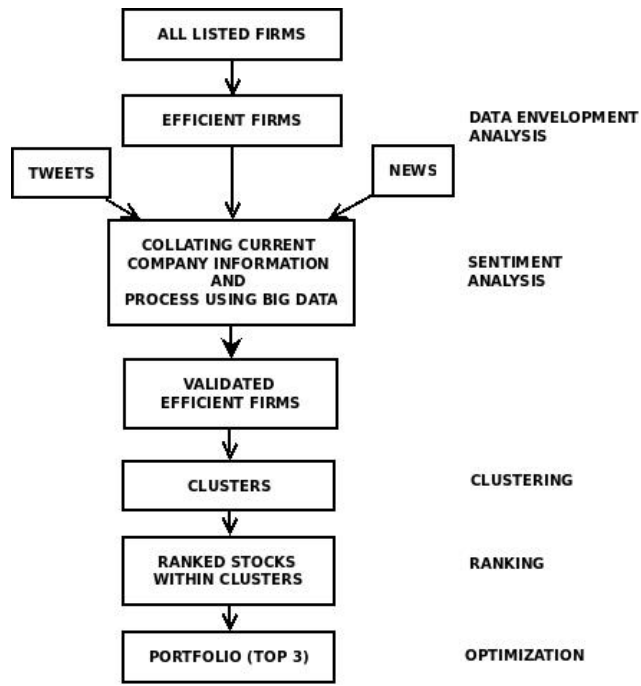

Figure 1: Framework for Portfolio Optimization

**2.1 Data Envelopment Analysis**

Data Envelopment Analysis (DEA) is a non-parametric linear programming that calculates the efficiency score of a Decision-Making Unit (DMU) based on a given set of inputs and outputs.

The DMUs with score 1 are considered to be efficient [7]. Apart from its applicability in the discipline of manufacturing, DEA can be used for stock selection. In this case, stocks form the DMU. Based on the previous studies [8, 9], four input parameters, namely, *total assets, total equity, cost of sales* and *operating expenses* and two output parameters, namely, *net sales and net income*, can be considered. These data can be obtained from standard databases like *Bloomberg*. The stocks with efficiency score 1 are considered for next stage.

## 2.2 Hadoop Framework for Sentiment Analysis

This stage involves processing of frequently generated unstructured data using Hadoop MapReduce. This step complements the previous stage. Events like election, change of management and announcement of dividends have an effect on the market sentiment, which is not captured using the quantitative analysis. As first step, online news articles and tweets of the efficient firms are retrieved. Tweets can be obtained through Twitter API but it is limited to 1500 tweets. The ease-of-use, scalability, and failover properties make Hadoop MapReduce a popular choice for processing big data efficiently [10]. Tweets and news articles are processed using text mining to obtain the positive and negative sentiments about the firm. Hadoop MapRaduce infrastructure quickens the distributed text mining process. Figure 2 shows the MapReduce framework for distributed text mining. As shown in the figure, the company tweets and news articles are distributed among different Map processors to produce intermediate data. This intermediate data is processed by the Reduce processors to give the aggregated result. The firms with positive sentiments are chosen for the next stage.

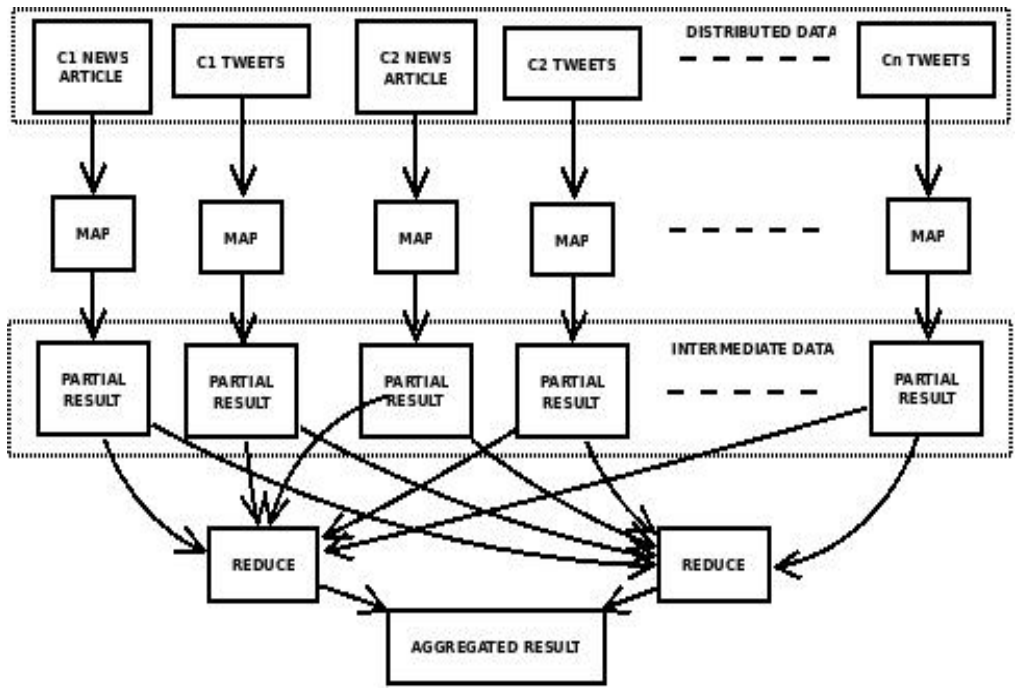

Figure 2: Hadoop framework for Distributed Text Mining

## 2.3 Stock Clustering

In this stage, the correlation coefficients of the returns of the stocks are calculated. The stocks are assigned to different clusters based on these correlation coefficients. The greater the number of clusters more is the diversification. The objective for number of clusters and quality of clusters is to maximize similarity within the cluster and to minimize the similarity between the clusters. Various clustering algorithms like k-means clustering or Louvian clustering [11] can be used. This process reduces the portfolio risk through diversification of stocks [12]. These resulting clusters can consist of firms with similar business activity or size in real sense.

## 2.4 Stock Ranking

The appropriate stocks from each cluster should be selected. The stocks in each cluster can be ranked using Artificial Neural Network (ANN) [13]. Till the previous stage, only the internal factors of the firms were considered. At this stage, external factors like Gross Domestic Product (GDP) growth rate and interest rate can be considered [12]. ANN is a model for information processing that consists of three layers: *input, hidden and output layer* respectively. The inputs for ANN can be GDP growth rate and interest rate and the outputs can be future return on investments. This results in ranking the stocks within a cluster.

## 2.5 Optimization

Previous stage leads to an assumption that the investor might choose the top stocks from each cluster. But the question that still remains is: How much to invest in each stock? Previous study [12] considered simple (equal) stock weighting method, a primitive method, to allocate the resources among the stocks. Hence the ranked stocks should be optimized to maximize returns and to minimize risk. Markowitz's mean-variance model can be used at this stage [6]. Various optimization heuristics like Particle Swarm Optimization (PSO), Ant Colony Optimization (ACO) and Genetic Algorithm (GA) can be used. The distribution of the stocks in a portfolio will be formed at the end of this stage. Top 3 performing portfolio will be suggested to the investor.

## 3. Expected Outcomes

Apart from letting the investors make an informed investment decision, this framework would be useful to the naïve investors as well. The expected outcome of the framework is to generate portfolio in conformance with the criteria of the investors: *minimize the risk* and *maximize the return*. The first stage of DEA would result in a group of the potential stocks for portfolio optimization. The current information related to those resulting companies is analyzed using text mining to obtain the sentiment about the company. Sentiment analysis gives the qualitative aspect of the firms. The stocks resulting from the previous stage are grouped into different categories using the correlation co-efficient of the returns. This leads to the diversification of the stocks. The stocks are ranked to select appropriate stocks from each group. At the end of the entire process, top three portfolio suggestions would be provided to the investors so that they can select one of those three.

## 4. Conclusion

The proposed framework tries to integrate both structured data from database (stock price, balance sheet data etc) and unstructured data from online news articles and tweets. Consideration of qualitative factors (Management of firms, etc.) along with quantitative factors (financial ratios) provides better alternatives for formation of portfolio. The assets are well-diversified using k-means clustering and ANN. Top three portfolio suggestions obtained using optimization heuristics gives flexibility to the investors to choose the appropriate assets suiting their risk profile. The proposed model can be applied to any stock market data. The utility of the framework is limited only to stock analysis and investments. The parameters suggested for DEA is based on the previous studies. Instead of variance as risk measure, other risk measures like Value-at-risk and downside risk measures can be used in the final stage of the framework.